\documentclass[sigplan,10pt]{acmart}
\AtBeginDocument{%
  }

\setcopyright{acmlicensed}
\copyrightyear{2027}
\acmYear{2027}
\acmDOI{XXXXXXX.XXXXXXX}
\acmConference[PPoPP '27]{The 32nd ACM SIGPLAN Annual Symposium on
  Principles and Practice of Parallel Programming}{TBD}{TBD}
\acmISBN{978-x-xxxx-xxxx-x/YY/MM}
\renewcommand\footnotetextcopyrightpermission[1]{}

\usepackage{booktabs}
\usepackage{enumitem}
\usepackage{subcaption}
\usepackage{xspace}
\usepackage{amsmath}
\usepackage{float}
\usepackage{tikz}
\usetikzlibrary{arrows.meta,positioning,calc,shapes.geometric,fit,backgrounds}
\definecolor{cprteal}{HTML}{2A9D8F}
\definecolor{cprblue}{HTML}{2563EB}
\definecolor{cprgold}{HTML}{D97706}
\definecolor{cprslate}{HTML}{334155}
\definecolor{cprbg}{HTML}{F5F0E8}
\definecolor{cprbbg}{HTML}{E8EDF2}
\usepackage{algorithm}
\usepackage[noend]{algpseudocode}
    \algrenewcommand\algorithmicindent{1.0em} 

\newcommand{\sys}{FaCTz\xspace}

\begin{document}

%\title{{\sys}: Fast Critical-Point and Topology-Aware Error-Bounded GPU Compression for Scientific Vector Fields}

\title{{\sys}: Fast Critical-Point and Topology-Aware GPU Compression for Scientific Vector Fields}

%% Authors anonymized for review.
% \author{Anonymous Author(s)}
% \affiliation{%
%   \institution{Submission \#XXX}
%   \country{}
% }

%% Authors
\author{%
{\large
Mingze Xia$^{*}$,
Yuxiao Li$^{\dagger}$,
Sheng Di$^{\ddagger}$,
Jiannan Tian$^{\star}$,
Baixi Sun$^{\ddagger}$
\\[2pt]
Boyi Zhang$^{\diamond}$,
Bei Wang$^{\circ}$,
Hanqi Guo$^{\dagger}$,
Xin Liang$^{*}$
}
\\[4pt]
{\normalsize
$^{*}$Oregon State University \quad
$^{\dagger}$The Ohio State University \quad
$^{\ddagger}$Argonne National Laboratory
\\[1pt]
$^{\star}$Oakland University \quad
$^{\diamond}$University of Kentucky \quad
$^{\circ}$University of Utah
}
}

\renewcommand{\shortauthors}{Xia et al.}

%% ===================================================================
%% ABSTRACT  (~0.25 page; 5-sentence structure)
%% ===================================================================
\begin{abstract}
Error-bounded lossy compression is essential for storing and transferring the vector-field data produced by large-scale scientific simulations. Although it enforces a user-specified error bound to limit numerical distortion, it does not preserve the field's \emph{topology}: small admissible perturbations can create or eliminate critical points on which downstream feature analysis depends. Existing GPU compressors achieve high throughput but are topology-agnostic, whereas the only compressor with provable critical-point preservation (cpSZ) runs on the CPU at throughput far below the data-generation rates of modern GPU-based systems. We observe that, although preserving critical points is inherently a coupled and sequential constraint, it can be reformulated into independent parallel tasks, either on a per-block basis or, speculatively, on a per-point basis. We present {\sys}, the first GPU-based error-bounded lossy compressor that guarantees critical-point preservation. {\sys} provides a block-wise mode optimized for throughput and a speculative per-point mode optimized for compression ratio. Across three vector-field datasets, {\sys} preserves every critical point while achieving throughput of up to 60\,GB/s, approximately two orders of magnitude (up to $\sim\!640\times$) faster than the multithreaded CPU implementation of cpSZ. Its speculative mode further improves the compression ratio by approximately a factor of two over the throughput-oriented mode.
\end{abstract}
% [Optional S5: availability / artifact statement once decided.]
% [Baseline (cuSZ/cuSZp/cuZFP) quality comparison to be folded in once run.]

%% --- CCS concepts (acmart requires these for papers over two pages) ---
%% NOTE for camera-ready: regenerate the official <ccs2012> XML block with the
%% ACM CCS tool (https://dl.acm.org/ccs) and paste it here inside a CCSXML
%% environment; the \ccsdesc lines below carry the same concepts and are what
%% the class typesets, but the ACM production system wants the XML too.


\maketitle

%% ===================================================================
%% S1 INTRODUCTION  (~1.25-1.5 pages)
%% Move 1 establish territory -> Move 2 find niche -> Move 3 occupy niche
%% ===================================================================
\section{Introduction}
\label{sec:intro}

Modern scientific simulations running on GPU-accelerated supercomputers generate data at rates that far exceed the available bandwidth for storage and transfer. A single high-resolution simulation of fluids, combustion, or plasma systems can routinely produce terabytes of vector-field data in a single run~\cite{kaneda2007dns,yeung2025gpu}. Even on leadership-class systems, transferring these datasets remains time-consuming, and the gap between computational throughput and I/O bandwidth continues to widen~\cite{kettimuthu2018petabyte,foster2017online,cappello2025practice}.

% A single high-resolution simulation of a fluid, combustion, or plasma
% system routinely emits vector-field snapshots on the order of terabytes per
% run; moving even a petabyte across leadership-class networks still takes on
% the order of a day~\cite{kettimuthu2018petabyte}, and the gap between compute
% throughput and I/O bandwidth continues to
% widen~\cite{foster2017online, cappello2025practice}. 

Lossless compression achieves only modest compression ratios on floating-point data, typically well below $2\times$~\cite{son2014data,lindstrom2017error}. In contrast, \emph{error-bounded lossy} compression guarantees that every reconstructed value remains within a user-specified error bound of the original while delivering substantially higher compression ratios~\cite{cappello2019usecases,di2025survey}.

A pointwise error bound, however, does not necessarily preserve the information that scientific users care about. For vector-field data, the quantities of interest are often \emph{topological}: the \emph{critical points} where the vector field vanishes, together with the flow structures anchored to them, form the foundation of feature extraction, vortex detection, and separation/attachment analysis~\cite{laramee2004state,laramee2007topology,heine2016survey,matsuoka2016new}. Even a small perturbation that satisfies the prescribed error bound can alter the sign pattern within a cell, thereby \emph{creating or destroying} a critical point. As a result, the reconstructed field satisfies the numerical error bound while encoding a different topology, potentially leading to incorrect scientific conclusions in downstream analyses. Consequently, although bounding the pointwise error is necessary, it is not sufficient for topology-aware analysis of vector fields. Preserving topology during compression is therefore essential to maintaining the integrity of subsequent scientific analyses.

This requirement exposes three fundamental gaps that no existing compressor addresses (\S\ref{sec:bg:motivation}). 

\begin{enumerate}[label=\textbf{G\arabic*.},ref=G\arabic*,leftmargin=2.2em,itemsep=3pt,topsep=3pt,parsep=0pt]

\item State-of-the-art GPU compressors~\cite{tian2020cusz,huang2023cuszp,huang2024cuszp2,lindstrom2014zfp,chen2025hpdr,liu2024cuszi,wu2025cuszhi} are topology-agnostic and can readily corrupt critical points. 

\item The only compressor that provably preserves critical points, cpSZ~\cite{liang2023cpsz}, achieves only $\sim\!0.1$--$0.2$\,GB/s even when using 128 CPU threads.

\item Critical-point preservation itself appears inherently difficult to parallelize because the admissible error at one point depends on the values of its neighbors.

\end{enumerate}
% \noindent

% \noindent\textbf{G2}:  

% \noindent\textbf{G3}: 

Our key insight is that critical-point preservation can be \emph{decomposed into independent parallel tasks} in two complementary ways, each targeting a different point in the throughput--compression-ratio tradeoff. First, partitioning the field into small blocks enables independent computation of a conservative error bound that preserves all critical points within each block, allowing the blocks to be compressed in parallel without cross-block dependencies. Second, tighter per-point error bounds are obtained through a hierarchical \emph{speculative} kernel that progressively enlarges each point's admissible error bound until the topological constraint is nearly violated, then retreats to the last valid bound. This approach trades some throughput for a substantially higher compression ratio. Together, these strategies transform an inherently sequential, dependency-laden constraint into one well suited to massively parallel GPU execution.

We embody this insight in {\sys} (\emph{Fast Critical-point and Topology-aware Compression}), the first GPU compressor for scientific vector fields that guarantees critical-point preservation under error-bounded lossy compression, thereby closing G1--G3. {\sys} provides two complementary operating modes: a \emph{block-wise} mode optimized for throughput and a \emph{speculative per-point} mode optimized for compression ratio. The main contributions are: 

\begin{enumerate}[label=\textbf{C\arabic*.},ref=C\arabic*,leftmargin=2.2em,itemsep=3pt,topsep=3pt,parsep=0pt]
\item We formulate critical-point preservation as a \emph{parallelizable} constraint, enabling the first GPU-based error-bounded compressor for vector fields with guaranteed critical-point preservation (\S\ref{sec:design}).

\item We instantiate this formulation in \sys through two complementary parallel strategies: a block-wise common-bound scheme and a hierarchical speculative per-point scheme, exposing a tunable throughput--compression-ratio tradeoff (\S\ref{sec:design}).

\item We develop a high-throughput CUDA implementation and present the key engineering techniques that enable its performance (\S\ref{sec:impl}).

\item We evaluate \sys against state-of-the-art GPU compressors (cuSZ, cuSZ-i, cuSZp, and cuZFP) and the CPU-based cpSZ on three vector-field datasets, comparing compression ratio, PSNR, compression and decompression throughput, and critical-point preservation (\S\ref{sec:eval}).
\end{enumerate}

%% ===================================================================
%% S2 BACKGROUND & MOTIVATION  (~1-1.5 pages)
%% ===================================================================
\section{Background and Motivation}
\label{sec:background}

\sys builds on three ingredients: the way error-bounded lossy compressors
discard information (\S\ref{sec:bg:ebc}), what defines a critical point in a 2D
vector field (\S\ref{sec:bg:cp}), and how cpSZ turns the two into a
preservation guarantee (\S\ref{sec:bg:cpsz}). We review them, then draw out the
observations that motivate a topology-preserving GPU compressor
(\S\ref{sec:bg:motivation}).

\subsection{Error-Bounded Lossy Compression}
\label{sec:bg:ebc}

Error-bounded lossy compressors~\cite{di2016sz, lindstrom2014zfp, liang2018error, zhao2021optimizing, ainsworth2019mgard, li2023sperr, liu2024high} keep every reconstructed value $\tilde{x}_i$
within a user-specified error bound $\epsilon$ of the original, $|\tilde{x}_i -
x_i|\le\epsilon$. \sys follows the dominant
\emph{prediction-based} design of SZ~\cite{di2016sz,
liang2018error, zhao2021optimizing}: a predictor (a Lorenzo or
interpolation stencil) estimates each value from already-decoded neighbors,
the residual is quantized into an integer bin sized by $\epsilon$ (the only
lossy step), and the codes are entropy-coded losslessly. Reconstructing each
value in place before predicting the next keeps encoder and decoder in
lockstep, so error never drifts. The guarantee is \emph{pointwise}: it bounds
each value's error but says nothing about relationships among values,
which is what topological features depend on.

\subsection{Critical Points in 2D Vector Fields}
\label{sec:bg:cp}

\begin{figure}[t]
  \centering
  \begin{tikzpicture}[
      >={Stealth[length=1.6mm]},
      vec/.style={->, semithick, cprblue},
      vtx/.style={circle, fill=cprslate, inner sep=0.9pt},
      cp/.style={star, star points=5, star point ratio=2.2, fill=cprgold,
                 draw=cprslate, inner sep=1.2pt},
      sub/.style={font=\footnotesize\bfseries, text=cprslate},
      mesh/.style={cprslate!35, thin},
      mulbl/.style={font=\footnotesize, text=cprslate, fill=white, inner sep=1.2pt},
      vlbl/.style={font=\scriptsize, text=cprblue!70!black, fill=white, inner sep=0.8pt},
      lead/.style={cprslate!55, thin},
    ]
    % ============ (a) zero falls INSIDE the highlighted cell ============
    \begin{scope}
      \foreach \j in {0,1,2}{\draw[mesh] (0,\j)--(2,\j);}
      \foreach \i in {0,1,2}{\draw[mesh] (\i,0)--(\i,2);}
      \draw[mesh] (0,0)--(1,1); \draw[mesh] (1,0)--(2,1);
      \draw[mesh] (0,1)--(1,2); \draw[mesh] (1,1)--(2,2);
      \begin{scope}[on background layer]
        \fill[cprteal!16] (0,0)--(1,0)--(1,1)--cycle;
      \end{scope}
      \draw[cprslate, semithick] (0,0)--(1,0)--(1,1)--cycle;
      \foreach \i in {0,1,2}{\foreach \j in {0,1,2}{\node[vtx] at (\i,\j){};}}
      \draw[vec] (0,0)--++(0.44,0.26);
      \draw[vec] (1,0)--++(-0.30,0.40);
      \draw[vec] (1,1)--++(-0.22,-0.46);
      \node[cp] at (0.7,0.4){};
      \node[mulbl] (mua) at (1.0,2.5){$\mu=(0.3,0.3,0.4)$};
      \draw[lead] (mua.south) -- (0.74,0.58);
      \node[vlbl, below=1pt] at (0,0){$(u_0,v_0)$};
      \node[vlbl, below=1pt] at (1,0){$(u_1,v_1)$};
      \node[vlbl, above right=-2pt] at (0.7,1.3){$(u_2,v_2)$};
      \node[sub] at (1.0,-0.62){(a) $0\le\mu_k\le1$};
    \end{scope}
    % ============ (b) zero falls OUTSIDE the highlighted cell ============
    \begin{scope}[xshift=3.6cm]
      \foreach \j in {0,1,2}{\draw[mesh] (0,\j)--(2,\j);}
      \foreach \i in {0,1,2}{\draw[mesh] (\i,0)--(\i,2);}
      \draw[mesh] (0,0)--(1,1); \draw[mesh] (1,0)--(2,1);
      \draw[mesh] (0,1)--(1,2); \draw[mesh] (1,1)--(2,2);
      \begin{scope}[on background layer]
        \fill[cprslate!8] (0,0)--(1,0)--(1,1)--cycle;
      \end{scope}
      \draw[cprslate, semithick] (0,0)--(1,0)--(1,1)--cycle;
      \foreach \i in {0,1,2}{\foreach \j in {0,1,2}{\node[vtx] at (\i,\j){};}}
      \draw[vec] (0,0)--++(0.5,0.16);
      \draw[vec] (1,0)--++(0.5,0.16);
      \draw[vec] (1,1)--++(0.5,0.16);
      \node[cp] at (1.5,0.9){};
      \node[mulbl] (mub) at (1.0,2.5){$\mu=(-0.5,0.6,0.9)$};
      \draw[lead] (mub.south) -- (1.44,0.98);
      \node[vlbl, below=1pt] at (0,0){$(u_0,v_0)$};
      \node[vlbl, below=1pt] at (1,0){$(u_1,v_1)$};
      \node[vlbl, above right=-2pt] at (0.7,1.3){$(u_2,v_2)$};
      \node[sub] at (1.0,-0.62){(b) some $\mu_k\notin[0,1]$};
    \end{scope}
  \end{tikzpicture}
  % \vspace{-2mm}
  \caption{A piecewise-linear cell in a triangular mesh contains a critical
  point iff the barycentric solution of Eq.~\eqref{eq:def} satisfies
  $0\le\mu_k\le1$ for all $k$. Left: a critical point exists; right: none. The
  right cell's linear field still vanishes at the $\star$, but a negative $\mu_k$ places
  that zero outside the cell. Therefore, the current cell contains no critical point, 
  and the neighboring cells must be examined separately.}
  \label{fig:cpexample}
  \Description{Two triangle cells with a vector arrow at each vertex. The left
  cell's vectors enclose an interior zero marked by a star, whose barycentric
  coordinates all lie in the unit interval. In the right cell the barycentric
  solution has a negative coordinate, so the star (the zero) sits outside the
  triangle and there is no interior critical point.}
\end{figure}
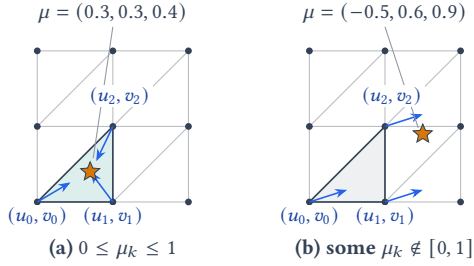

We consider 2D vector fields sampled at the vertices of a simplicial
(triangular) mesh, where each vertex carries a two-component vector $(u,v)$
and the field is reconstructed by piecewise-linear interpolation over each
cell. A \emph{critical point} is a location where the vector field vanishes (i.e., the interpolated $u$ and $v$ are both 0).
Under the piecewise-linear assumption, the critical point of a cell is found
by solving a linear system in barycentric coordinates:
\begin{equation}
\label{eq:def}
    \begin{bmatrix}
    u_0 & u_1 & u_2\\
    v_0 & v_1 & v_2
    \end{bmatrix}
    \begin{bmatrix}
    \mu_0\\
    \mu_1\\
    \mu_2
    \end{bmatrix} = \mathbf{0} \text{\ and\ } \mu_0 + \mu_1 + \mu_2 = 1,
\end{equation}
where $(\mu_0, \mu_1, \mu_2)$ are the barycentric coordinates and $(u_i,
v_i)$ are the vector components at the three vertices of the cell.
By Cramer's rule the barycentric solution is
\begin{equation}\label{eq:bary}
  \mu_k = \frac{m_k}{m},\enspace
  m_k = u_{k+1}v_{k+2} - u_{k+2}v_{k+1},\enspace
  m = m_0 + m_1 + m_2,
\end{equation}
with indices taken modulo $3$; each $m_k$ is thus the
$2\times2$ determinant of the vectors at the other two vertices.
A critical
point exists within the cell if and only if $0 \le \mu_k \le 1$ for all $k
\in \{0,1,2\}$, i.e.\ every $m_k$ shares the sign of $m$ with
$|m_k| \le |m|$. Figure~\ref{fig:cpexample} illustrates cells with and without
a critical point.

Beyond mere existence, each critical point has a \emph{type} (source, sink,
saddle) determined by the eigenvalues of the Jacobian
of the linear field over the cell~\cite{helman1991topology, theisel2008topological}.
Concretely, the eigenvalues of the Jacobian $J$ solve
the characteristic equation $\lambda^2 - \operatorname{tr}(J)\,\lambda +
\det(J) = 0$, so with discriminant $\Delta = \operatorname{tr}^2(J) -
4\det(J)$ the type is fixed by three signs: $\det(J) < 0$ gives a
\emph{saddle}; $\det(J) > 0$ gives a \emph{source} when $\operatorname{tr}(J) >
0$ and a \emph{sink} when $\operatorname{tr}(J) < 0$, which spirals when
$\Delta < 0$ (complex eigenvalues) and is a plain node when $\Delta \ge 0$.
Preserving a point's type thus amounts to preserving the signs of
$\det(J)$, $\operatorname{tr}(J)$, and $\Delta$.
Type matters downstream: feature extraction, vortex detection, and
separatrix tracing all classify and connect critical points by type, so a
type change is as damaging as a spurious or missing point.

Accordingly, we say a compressor \textbf{preserves critical points} when the
reconstructed field satisfies all of the following with respect to the
original: \textbf{(P1)} every original critical point survives, in the same
cell (no loss of position); \textbf{(P2)} no new critical point appears where
the original had none (no spurious points); and \textbf{(P3)} every preserved
critical point retains
its type (e.g., a source never becomes a saddle). \sys enforces P1--P3
jointly, on top of the pointwise error bound of
\S\ref{sec:bg:ebc}.

\subsection{Critical-Point-Preserving Lossy Compression}
\label{sec:bg:cpsz}

cpSZ~\cite{liang2020toward, liang2023cpsz} is the reference method for P1--P3 and the starting
point for \sys. It keeps the SZ pipeline of \S\ref{sec:bg:ebc} and adds one
stage in front of it: before a vertex is quantized, cpSZ derives from the sign
conditions of Eq.~\eqref{eq:def} a \emph{safe} error bound, the sufficient
but not necessary bound on the perturbation of that vertex under which no
incident triangle can change its critical-point signature. Quantization then runs under the per-vertex minimum
of that bound instead of the user's $\epsilon$, so preservation costs
compression ratio rather than correctness.

cpSZ derives the bound in two ways, which differ in whether the derivation is
coupled to compression. The \emph{decoupled} (offline) variant makes a
separate first pass that computes every vertex's bound from the
\emph{original} field, then compresses in a second pass; the bounds are
independent of one another, but they must be conservative enough to hold
whatever the neighbors decompress to. The \emph{coupled} (online) variant
folds the derivation into the compression sweep, deriving each vertex's bound
against neighbors that have already been reconstructed. That is tighter
and compresses better, but it makes the bound at one vertex depend on
decisions at earlier ones, which serializes the sweep. \sys keeps this
distinction and resolves the parallelism obstacle in each: its block-wise mode
follows the decoupled route (\S\ref{sec:design:block}) and its speculative
mode the coupled one (\S\ref{sec:design:online}).

\subsection{Motivating Observations}
\label{sec:bg:motivation}

\begin{figure}[t]
  \centering
  \includegraphics[width=\columnwidth, trim={0 4ex 0 0}]{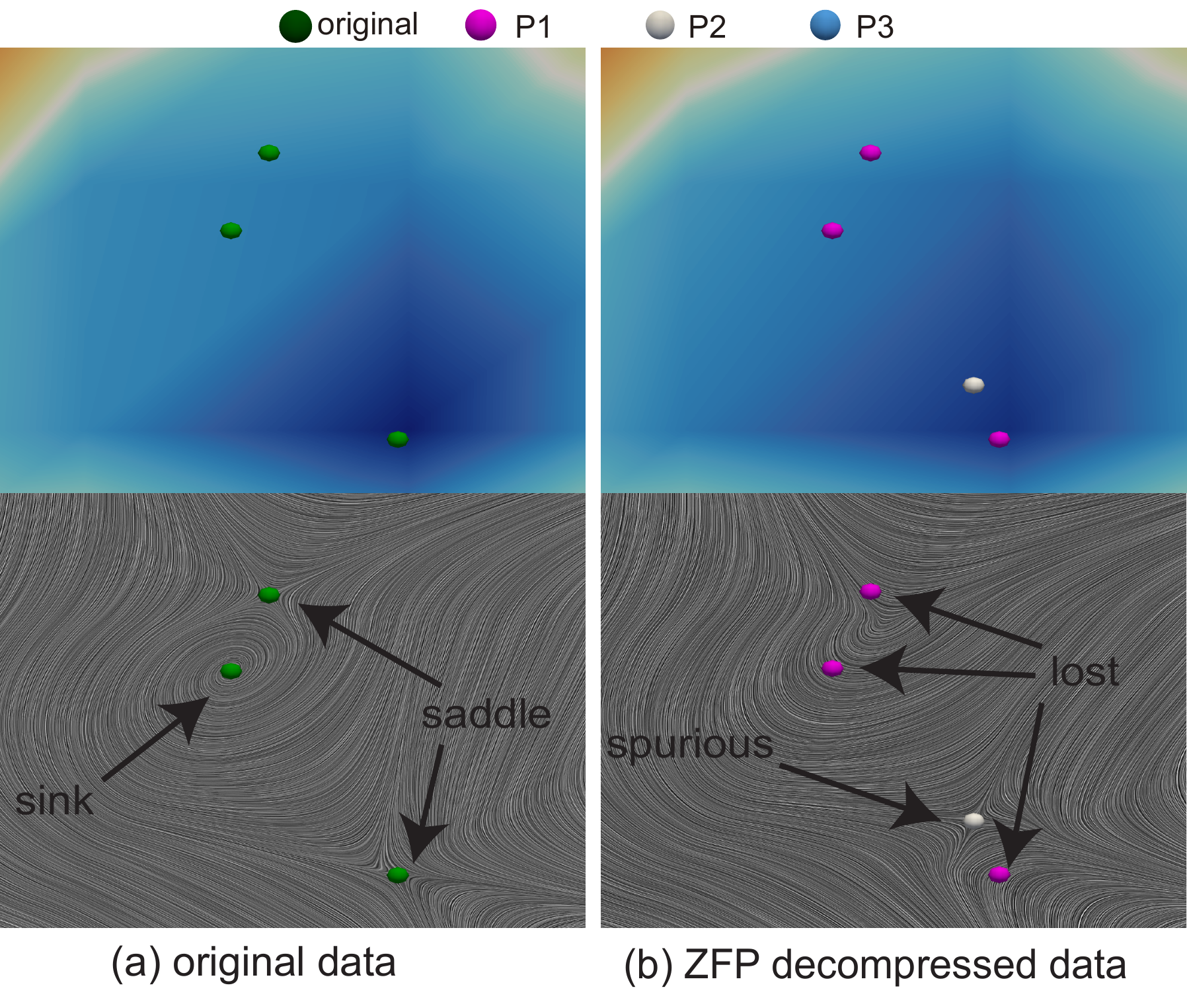}
  \caption{A pointwise error bound does not protect topology. (a)~Original field
  versus (b)~its ZFP reconstructed data. \emph{Top:}
  vector-magnitude rendering with critical points marked; \emph{Bottom:}
  line-integral-convolution (LIC) view of the flow. The magnitude fields look
  nearly identical, yet the reconstruction has \emph{lost} three original
  critical points and introduced a \emph{spurious} one, which corrupts the
  topology despite high pointwise fidelity (O1).}
  \label{fig:distortion}
  \Description{Original and ZFP-reconstructed vector field side by side, shown
  as vector-magnitude renderings with critical points marked and as
  line-integral-convolution flow views; the magnitude fields look nearly
  identical, but the reconstruction has lost three critical points and gained
  a spurious one.}
\end{figure}

% Each observation is backed by a concrete measurement and leads directly to a
% design implication.

\begin{description}[style=unboxed,leftmargin=0pt,itemsep=3pt,topsep=3pt,parsep=0pt]
  \item[\textbf{O1: Topology-agnostic GPU compressors corrupt critical points,
  even at tight error bounds.}]
  Because compressors such as cuSZ, cuSZp, and cuZFP quantize each value
  independently of any topological constraint, admissible per-value
  perturbations flip the sign structure of Eq.~\eqref{eq:def} and thereby
  violate P1--P3 (Figure~\ref{fig:distortion}). At the error
  bounds that yield useful compression they corrupt topology, losing and
  spuriously creating critical points (\S\ref{sec:eval:cp}); recovering
  it (when possible at all) demands a near-lossless bound at which the
  compression ratio collapses to barely above $1\times$
  (Table~\ref{tab:cpcmp}). High pointwise fidelity thus coexists with topological
  corruption unless compression is all but abandoned.
  \emph{Implication:} a pointwise error bound is necessary but not sufficient;
  the topological constraint must be built into the compressor.

  \item[\textbf{O2: The only critical-point-preserving compressor is CPU-bound.}]
  cpSZ provably preserves critical points but runs on the CPU at only
  $\sim\!0.1$--$0.2$\,GB/s even when parallelized across $128$ threads
  (\S\ref{sec:eval:cpsz}). GPU simulations emit vector-field
  snapshots orders of magnitude
  faster, so inserting cpSZ into the pipeline turns compression into the
  bottleneck rather than a remedy.
  \emph{Implication:} topology-preserving compression must itself run on the
  GPU, at a throughput commensurate with data generation.

  \item[\textbf{O3: No single operating point serves every deployment.}]
  Whether storage footprint or wall-clock time is the binding constraint
  varies across workflows. An in-situ pipeline racing alongside the simulation
  prioritizes throughput, whereas archival storage and wide-area data transfer
  prioritize compression ratio. A topology-preserving compressor should
  therefore expose a tunable trade-off rather than commit to a single
  operating point.
  \emph{Design implication:} \sys provides two configurations: a
  throughput-oriented block-wise mode and a ratio-oriented speculative mode
  (\S\ref{sec:design}).
\end{description}

%% ===================================================================
%% S3 DESIGN  (~3 pages)
%% ===================================================================
\section{Design}
\label{sec:design}

This section presents \sys's design: the two-mode architecture and shared
backbone (\S\ref{sec:design:overview}), the critical-point signature both modes
enforce (\S\ref{sec:design:signature}), and the two ways \sys turns it into a
safe error bound, an analytic block-wise mode (\S\ref{sec:design:block}) and a
speculative per-point mode (\S\ref{sec:design:online}).

\subsection{Overview}
\label{sec:design:overview}

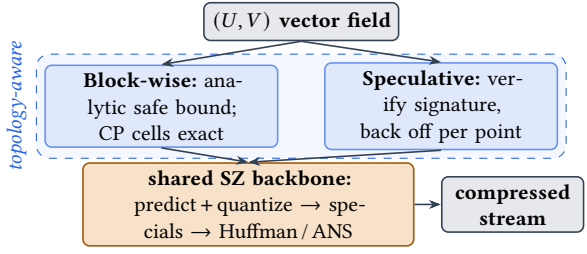
\begin{figure}[t]
  \centering
  % All text \footnotesize (8pt at the 10pt base), the CFP floor for figures.
  \begin{tikzpicture}[
      >={Stealth[length=1.6mm]},
      box/.style={rounded corners=2.5pt, draw=cprslate, semithick,
                  align=center, font=\footnotesize, inner sep=2.5pt},
      io/.style={box, fill=cprslate!12, font=\footnotesize\bfseries,
                 minimum height=5mm},
      fe/.style={box, fill=cprblue!14, draw=cprblue!70, text width=29mm,
                 minimum height=8mm},
      bb/.style={box, fill=cprgold!22, draw=cprgold!75!black, text width=42mm,
                 minimum height=8mm},
      flow/.style={->, semithick, cprslate},
      band/.style={rounded corners=5pt, dashed, inner sep=3pt},
    ]
    \node[io, text width=24mm] (in) at (0,0) {$(U,V)$ vector field};
    \node[fe] (feA) at (-1.85,-1.12)
      {\textbf{Block-wise:} analytic safe bound; CP cells exact};
    \node[fe] (feB) at ( 1.85,-1.12)
      {\textbf{Speculative:} verify signature, back off per point};
    \node[bb] (bbk) at (-0.7,-2.42)
      {\textbf{shared SZ backbone:} predict\,$+$\,quantize $\to$ specials
       $\to$ Huffman\,/\,ANS};
    \node[io, text width=17mm, anchor=west] (out) at ($(bbk.east)+(3.5mm,0)$)
      {compressed stream};

    \draw[flow] (in.south) -- (feA.north);
    \draw[flow] (in.south) -- (feB.north);
    \draw[flow] (feA.south) -- (bbk.north);
    \draw[flow] (feB.south) -- (bbk.north);
    \draw[flow] (bbk.east) -- (out.west);

    \begin{scope}[on background layer]
      \node[band, draw=cprblue, fill=cprblue!5, fit=(feA)(feB),
        label={[rotate=90,anchor=south,font=\footnotesize\itshape,text=cprblue]%
               left:topology-aware}] (fg) {};
    \end{scope}
  \end{tikzpicture}
  % \vspace{-1mm}
  \caption{\sys pipeline: a mode-specific, topology-aware frontend
  (block-wise or speculative) precedes the shared SZ backbone, so the
  critical-point check happens before entropy coding.}
  \label{fig:pipeline}
  \Description{Block diagram of the \sys pipeline: the input field forks into the
  block-wise and speculative frontends, which feed a shared SZ backbone and
  entropy coder that produce the compressed stream.}
\end{figure}

\sys reuses a standard SZ-style backbone (prediction, error-controlled
quantization with in-place reconstruction, special-value collection, and
lossless entropy coding) but replaces its frontend with a \emph{mode-specific},
\emph{topology-aware error-control} stage that decides, before any value is
committed, how much error each vertex may absorb without altering the field's
critical points (Figure~\ref{fig:pipeline}). Whereas SZ-style
backbones entropy-code with Huffman~\cite{huffman1952}, we additionally integrate an optional ANS
coder (via nvCOMP) for higher throughput (\S\ref{sec:eval:cfgthrpt}).

The two configurations do \emph{not} share a frontend; they differ in
when and how to resolve the topological constraint, which
places them at opposite ends of the speed-ratio spectrum. The
\textbf{block-wise} mode (\S\ref{sec:design:block}) resolves it
\emph{analytically and ahead of time}, deriving a conservative safe bound per
vertex under which an ordinary Lorenzo predictor compresses; the
\textbf{speculative per-point} mode (\S\ref{sec:design:online}) resolves it
\emph{online}, verifying at each vertex that the candidate reconstruction
keeps every signature intact and backing off the bound only on failure. Both
check critical points before entropy coding and provide a deterministic
preservation guarantee (\S\ref{sec:design:correctness}).

\subsection{The Critical-Point Signature}
\label{sec:design:signature}

Both modes rest on one primitive: deciding whether perturbing the vertex
vectors changes the \emph{critical-point signature} of a triangle. For a cell
with vertex vectors $(u_i,v_i)$, the signature records (i) whether a critical
point lies inside---i.e., whether the barycentric solution of
Eq.~\eqref{eq:def} satisfies $0\le\mu_k\le1$---and (ii) if so, its type, given
by the eigenstructure of the cell's Jacobian. Preserving P1--P3
(\S\ref{sec:bg:cp}) over the whole field is exactly preserving the signature
of \emph{every} triangle.

On a regular grid triangulated in the usual way, an interior vertex is shared
by up to six triangles, so the error admissible at a vertex is coupled to all
of them: a change is safe only if it preserves the signature of every
incident cell. This many-to-one coupling (one vertex, up to six constraints) is why
critical-point preservation resists naive parallelization, and the two modes
handle it differently.

Evaluating a signature exactly needs the barycentric solution of
Eq.~\eqref{eq:def} and a Jacobian eigen-classification, most robustly in
double precision, costly on GPUs whose FP64 throughput is a fraction of FP32's.
\sys therefore tests each triangle in two tiers: a single-precision interval
filter that conservatively proves the field cannot vanish in the
triangle (so its signature cannot change) settles the common case, and only
the few ambiguous triangles fall through to the exact test. Because most
triangles lie far from a critical point, this filter removes the bulk of the
double-precision work; it is what makes the per-point verification of
\S\ref{sec:design:online} affordable.

\subsection{Block-Wise Mode: Analytic Safe Bounds}
\label{sec:design:block}

The block-wise mode computes, ahead of prediction, a per-vertex error bound
that is guaranteed safe, in four steps:
\begin{enumerate}
  \item \textbf{Per-triangle safe bound.} For each triangle we derive a
  conservative \emph{relative} bound from the sign conditions of
  Eq.~\eqref{eq:def}, following the critical-point-preserving error analysis of
  Liang et al.~\cite{liang2023cpsz}: a closed-form \emph{sufficient} bound (not
  a searched, and not necessarily the maximal feasible, value) under which no
  admissible perturbation can flip the cell's signature.
  \item \textbf{Per-vertex minimum.} Each vertex takes the minimum bound over
  its (up to six) incident triangles, discharging the coupling of
  \S\ref{sec:design:signature}.
  \item \textbf{Absolute bound and quantization.} The relative bound is scaled
  by $|U|$ and $|V|$ to yield separate absolute bounds for the two
  components, discretized into integer error-bound IDs.
  \item \textbf{Tile common bound.} Within each $32\times32$ vertex tile we
  take the minimum ID separately for $U$ and $V$, yielding two per-tile common
  bounds $\mathrm{eb}_U$ and $\mathrm{eb}_V$ (the components are \emph{not}
  forced to share one bound). A single common bound per tile lets the Lorenzo
  predictor and quantizer run without per-vertex bound bookkeeping.
\end{enumerate}
Step~1 is derived as follows. Each sign condition of
Eq.~\eqref{eq:def} is the sign of a $2\times2$ determinant $D$ of the cell's
vector components (one of the barycentric numerators $m_k$ or the orientation
$m$ of Eq.~\eqref{eq:bary}), a signed sum of degree-two products of the values.
Writing
$D=P_{+}-P_{-}$, where $P_{+}$ and $P_{-}$ collect the magnitudes of its
positive and negative product terms, a value-range-relative perturbation of
size $\xi$ keeps every value within $v(1\pm\xi)$ and hence moves each degree-two
product by at most a factor of $(1\pm\xi)^2$; the sign of $D$ cannot flip while
$P_{+}(1-\xi)^2 \ge P_{-}(1+\xi)^2$, yielding the closed-form safe bound
\begin{equation}
\label{eq:eb}
  \xi_{\mathrm{safe}}(D)=\frac{\bigl|\sqrt{P_{+}}-\sqrt{P_{-}}\bigr|}
                              {\sqrt{P_{+}}+\sqrt{P_{-}}} .
\end{equation}
The per-triangle bound of step~1 is the minimum of $\xi_{\mathrm{safe}}$ over
the determinant-sign conditions that keep the cell critical-point-free~\cite{liang2023cpsz}; the
$\sqrt{\cdot}$ arises because each determinant term is a product of two
perturbed values.
Triangles that already contain a critical point (or are degenerate) return a
safe relative bound of zero, so all three of their vertices (both
components) receive error-bound ID $0$ and are recorded in a zero-EB
mask/value side stream that restores their original values on decompression.
A positive analytic bound therefore protects every triangle that has no
critical point (P2), and exact storage secures P1 and P3.

Because the per-vertex bound is derived before tiling and the tile minimum only
ever shrinks it, correctness does not depend on triangles staying within a
tile. Bound derivation is thus a fixed-stencil pass with no cross-tile
dependence, making the block-wise mode embarrassingly parallel
(\S\ref{sec:impl} gives the CUDA patch layout).

\subsection{Speculative Per-Point Mode: Parallel Verify-and-Back-Off}
\label{sec:design:online}

The speculative mode keeps the user's full error bound wherever topology
allows and tightens it only where a critical point would otherwise change,
which is why it reaches a higher ratio. \sys breaks the sequential dependence
of cpSZ's coupled variant (\S\ref{sec:bg:cpsz}) with a \emph{dyadic
multiresolution} schedule that exposes wide parallelism at every level while
keeping the encoder and decoder in lockstep.

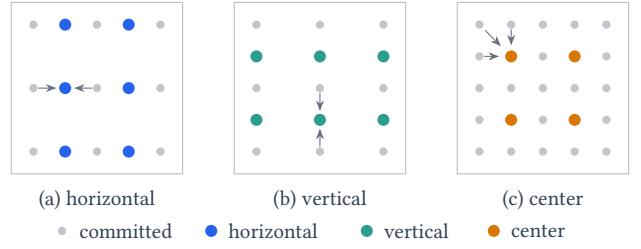
\begin{figure}[t]
  \centering
  % All text \footnotesize (8pt at the 10pt base), the CFP floor for figures.
  \begin{tikzpicture}[x=0.42cm,y=0.42cm,
      com/.style={circle,fill=cprslate!30,inner sep=1.1pt},
      hor/.style={circle,fill=cprblue,inner sep=1.6pt},
      ver/.style={circle,fill=cprteal,inner sep=1.6pt},
      cen/.style={circle,fill=cprgold,inner sep=1.6pt},
      ar/.style={->,>={Stealth[length=1.4mm]},cprslate!75,thin},
      sub/.style={font=\footnotesize,text=cprslate,anchor=north},
      leg/.style={font=\footnotesize,text=cprslate,anchor=west},
      frame/.style={cprslate!35,thin},
    ]
    % ---- (a) horizontal: targets between left/right committed neighbors ----
    \begin{scope}
      \draw[frame] (-0.7,-0.7) rectangle (4.7,4.7);
      \foreach \y in {0,2,4}{
        \foreach \x in {0,2,4}{\node[com] at (\x,\y){};}
        \foreach \x in {1,3}{\node[hor] at (\x,\y){};}}
      \draw[ar] (0.15,2) -- (0.72,2);
      \draw[ar] (1.85,2) -- (1.28,2);
      \node[sub] at (2,-0.9) {(a) horizontal};
    \end{scope}
    % ---- (b) vertical: targets between top/bottom committed neighbors ----
    \begin{scope}[xshift=2.95cm]
      \draw[frame] (-0.7,-0.7) rectangle (4.7,4.7);
      \foreach \x in {0,2,4}{
        \foreach \y in {0,2,4}{\node[com] at (\x,\y){};}
        \foreach \y in {1,3}{\node[ver] at (\x,\y){};}}
      \draw[ar] (2,0.15) -- (2,0.72);
      \draw[ar] (2,1.85) -- (2,1.28);
      \node[sub] at (2,-0.9) {(b) vertical};
    \end{scope}
    % ---- (c) center: targets read committed H/V/anchor neighbors ----
    \begin{scope}[xshift=5.9cm]
      \draw[frame] (-0.7,-0.7) rectangle (4.7,4.7);
      \foreach \x in {0,...,4}{\foreach \y in {0,...,4}{
        \ifodd\x \ifodd\y \node[cen] at (\x,\y){};
                 \else \node[com] at (\x,\y){}; \fi
        \else \node[com] at (\x,\y){}; \fi}}
      \draw[ar] (1,3.85) -- (1,3.35);     % from top
      \draw[ar] (0.15,3) -- (0.62,3);     % from left
      \draw[ar] (0.18,3.82) -- (0.68,3.32);% from top-left
      \node[sub] at (2,-0.9) {(c) center};
    \end{scope}
    % ---- legend: one row under the three panels ----
    \begin{scope}[yshift=-1.05cm]
      \node[com]  at (0.9,0){};  \node[leg] at (1.15,0){committed};
      \node[hor]  at (5.6,0){};  \node[leg] at (5.85,0){horizontal};
      \node[ver]  at (10.5,0){}; \node[leg] at (10.75,0){vertical};
      \node[cen]  at (14.5,0){}; \node[leg] at (14.75,0){center};
    \end{scope}
  \end{tikzpicture}
  % \vspace{-3mm}
  \caption{\sys-S parallel schedule for one refinement level: (a) horizontal
  targets, predictor $\tfrac12(x_L{+}x_R)$, and (b) vertical,
  $\tfrac12(x_T{+}x_B)$, are mutually independent; (c) center,
  $x_T{+}x_L{-}x_{TL}$, then reads both. Gray vertices are already committed;
  % arrows show one targets stencil.}
  arrows indicate stencil.}
  \label{fig:schedule}
  \Description{Three dot-grid panels: in the horizontal phase, targets sit
  between committed left and right neighbors; in the vertical phase, between
  committed top and bottom neighbors; in the center phase, each target reads
  its committed top, left, and top-left neighbors. A legend below names the
  four dot colors.}
\end{figure}

\noindent\textbf{Multiresolution schedule.} Over the same mesh (no downsampled
copies are built), an \emph{anchor} kernel first stores the vertices on a
stride-$s_0$ Cartesian lattice losslessly (default $s_0=16$); these anchors are
never predicted and serve as exact starting points, with a sentinel
eliding the degenerate $(U,V)=(0,0)$ case. Prediction then densifies the grid over
$\log_2 s_0$ levels, halving the step at each level down to $1$. Each level fills
the vertices lying halfway between the previous level's points in three
ordered \emph{phases} (Figure~\ref{fig:schedule}): \emph{horizontal} vertices,
predicted from the two committed neighbors at $\pm\text{step}$ along the row
($\hat{x}=\tfrac12(x_L+x_R)$); then \emph{vertical} vertices
($\hat{x}=\tfrac12(x_T+x_B)$); then \emph{center} vertices, from a 2D Lorenzo
stencil ($\hat{x}=x_T+x_L-x_{TL}$ at offset $-\text{step}$). One anchor kernel
plus three phases per level is $1+3\log_2 s_0$ kernel launches ($13$ at the
default $s_0{=}16$).

\noindent\textbf{Where the parallelism comes from.} Within a single phase,
every target vertex is spaced $2\times\text{step}$ from the next and reads only
\emph{already-committed} neighbors (anchors or coarser-level points), never
another target of the same phase. No two threads in a phase therefore touch a
shared triangle, so the phase is embarrassingly parallel: \sys maps one thread
to each target vertex and keeps hundreds of thousands of vertices in flight
(Figure~\ref{fig:schedule}). The phases themselves are \emph{not} parallel with
one another: the horizontal and vertical phases each read only the coarser
level, but the center phase reads the horizontal and vertical vertices it sits
between, so it must run after both. \sys therefore issues the three phases as
three ordered kernel launches per level (horizontal, vertical, then center),
and the levels coarse-to-fine. Each committed value is written in place, so the next level reads the
\emph{decompressed} value and the decompressor, replaying the identical
13-launch schedule, reconstructs bit-for-bit.

Algorithm~\ref{alg:spec} summarizes the end-to-end pipeline. At each target
vertex the mode runs deferred-commit speculation: it forms candidate
reconstructions for $U$ and $V$ at the current bound, then checks the
signatures of all (up to six) incident triangles.
A triangle's signature encodes both whether it contains a
critical point and, if so, its type: the signs of the barycentric numerators
together with those of $\det(J)$, $\operatorname{tr}(J)$, and $\Delta$
(\S\ref{sec:bg:cp}). Unlike the block-wise mode, the speculative mode preserves
these type signs directly through an analogous safe bound. According to~\cite{liang2023cpsz}, the determinant and
trace signs give a bound of the Eq.~\eqref{eq:eb} form, but the discriminant
$\Delta$ is \emph{quadratic} in the perturbation $\xi$; the same sign-margin
argument then yields the safe bound as a root of that quadratic,
\begin{equation}
\label{eq:ebtype}
  \xi_{\mathrm{type}} = \left( \sqrt{L^{2}-4Q\Delta}-L \right) /{2Q},
\end{equation}
where $L$ and $Q$ are the first- and second-order sensitivities of $\Delta$ to
$\xi$ (recovering the linear bound $|\Delta|/L$ of Eq.~\eqref{eq:eb} as
$Q\!\to\!0$). This lets a critical-point-bearing cell keep a positive bound
instead of being stored exactly.
If every signature is unchanged, it commits the candidate
reconstruction and quantization codes. If any check fails, it decrements the
error-bound ID of both components and retries, down to $\mathrm{eb}=0$. Bounds
live on a logarithmic grid, so the ID is the log of the bound it encodes and
one decrement tightens that bound by a fixed factor
(Algorithm~\ref{alg:spec}). If even the exact candidate fails (a
degenerate configuration), the vertex's $(U,V)$ pair is stored as a raw
special value. Committing a candidate is the first write-back, so a failed
attempt costs nothing to undo: the retry is a purely local, deferred commit
rather than a block-wide rollback.

\begin{algorithm}
\small
\caption{\sys-S: end-to-end speculative compression}
\label{alg:spec}
\begin{algorithmic}[1]
\Require field $(U,V)$ on device; relative bound $\tau$; anchor stride $s_0$
\Ensure compressed stream (device-resident)
\State $\epsilon \gets \tau\cdot\mathrm{range}(U,V)$
       \Comment{absolute bound}
\State $\mathrm{eb}(\mathrm{id}) \gets \mathrm{eb}_{\min}\,\beta^{\,\mathrm{id}}$;\
       $\mathrm{id}_{\max}\gets\bigl\lfloor\log_\beta(\epsilon/\mathrm{eb}_{\min})\bigr\rfloor$
       \Comment{bounds on a logarithmic grid, so the stored $\mathrm{id}$
                is the log of the bound}
\State store every $s_0$-th vertex exactly (anchors)
       \Comment{one kernel}
\For{$\text{step}=s_0/2,\,s_0/4,\,\dots,\,1$}
       \Comment{$\log_2 s_0$ levels}
  \For{phase $\in$ \{horizontal, vertical, center\}}
       \Comment{one kernel each; launch = global barrier}
    \ForAll{target vertices of the phase \textbf{in parallel}}
      \State $\sigma_T \gets \Call{Signature}{T}$ for each incident triangle
             $T$ (\S\ref{sec:design:signature})
      \For{$\mathrm{id}=\mathrm{id}_{\max}$ \textbf{down to} $0$}
             \Comment{each step down tightens the bound by $\beta$}
        \State candidate $(\tilde u,\tilde v)\gets$ predict from committed
               neighbors ($\pm$step), quantize at bound $\mathrm{eb}(\mathrm{id})$
        \If{$\Call{Signature}{T}=\sigma_T$ for every incident $T$}
          \State commit $(\tilde u,\tilde v)$, codes $(q_u,q_v)$, and
                 $\mathrm{id}$; \textbf{break}
        \EndIf
      \EndFor
      \State \textbf{if} nothing committed: mark vertex raw (exact $(u,v)$)
    \EndFor
  \EndFor
\EndFor
\State compact raw/anchor values: bit-mask $\to$ prefix scan $\to$ side stream
\State entropy-code the four streams $q_U,q_V$ and
       $\mathrm{id}_U,\mathrm{id}_V$ (ANS)
\State \Return coded blobs $+$ raw side stream $+$ header
\end{algorithmic}
\end{algorithm}

% \begin{figure}[!htbp]
% \begin{lstlisting}[style=algostyle,
%   caption={\sys-S: end-to-end speculative compression},
%   label={lst:spec}]
% Require: field $(U,V)$ on device; relative bound $\tau$; anchor stride $s_0$
% Ensure : compressed stream (device-resident)

% $\epsilon \gets \tau\cdot\mathrm{range}(U,V)$                 // absolute bound
% store every $s_0$-th vertex exactly (anchors)                 // one kernel

% for step = $s_0/2,\,s_0/4,\,\dots,\,1$:                       // $\log_2 s_0$ levels
%   for phase in {horizontal, vertical, center}:                // one kernel each; launch = global barrier
%     for all target vertices of the phase in parallel:
%       $\sigma_T \gets \textsc{Signature}(T)$ for each incident triangle $T$   (*\S\ref{sec:design:signature}*)
%       for $\mathrm{id} = \mathrm{id}_{\max}$ down to $0$:
%         candidate $(\tilde u,\tilde v) \gets$ predict from committed
%             neighbors ($\pm$step), quantize at bound $\mathrm{id}$
%         if $\textsc{Signature}(T) = \sigma_T$ for every incident $T$:
%           commit $(\tilde u,\tilde v)$, codes $(q_u,q_v)$, and $\mathrm{id}$
%           break
%       if nothing committed: mark vertex raw (exact $(u,v)$)

% compact raw/anchor values: bit-mask $\to$ prefix scan $\to$ side stream
% entropy-code the four streams $q_U,q_V$ and $\mathrm{id}_U,\mathrm{id}_V$ (ANS)
% return coded blobs $+$ raw side stream $+$ header
% \end{lstlisting}
% \end{figure}

This dependency-ordered schedule makes the speculative mode slower than
block-wise, but by letting most vertices keep the full requested bound
(\S\ref{sec:design:correctness} proves preservation) it reaches a
substantially higher ratio.

% \emph{Alternative considered.} A block-level rollback (speculate an entire
% tile, then undo on any violation) would simplify scheduling but wastes work
% proportional to tile size on every violation and reintroduces a bulk
% dependence, especially at tile boundaries; per-vertex deferred commit localizes both the check and the cost.

\subsection{Correctness}
\label{sec:design:correctness}

Let $\sigma(T;x)$ denote the critical-point signature of triangle $T$
evaluated on a field $x$ (\S\ref{sec:design:signature}). By construction,
P1--P3 hold iff $\sigma(T;\tilde{x})=\sigma(T;x)$ for \emph{every} triangle
$T$, where $\tilde{x}$ is the decompressed data; this is what both modes prove.

\emph{Block-wise mode.} Eq.~\eqref{eq:eb} is a \emph{sufficient} bound: if
every vertex of a critical-point-free triangle $T$ carries relative error at
most $\xi_{\mathrm{safe}}(D)$ for each determinant $D$ in $T$'s signature,
then $P_{+}(1-\xi)^2 \ge P_{-}(1+\xi)^2$ still holds, no determinant sign can
flip, and $\sigma(T;\tilde{x})=\sigma(T;x)$. The mode enforces this
hypothesis: each vertex is quantized under the minimum of
$\xi_{\mathrm{safe}}$ over its incident triangles, and the per-tile reduction
only tightens that bound, so every critical-point-free triangle keeps its
signature (P2); critical-point-bearing and degenerate triangles are stored
exactly ($\mathrm{eb}=0$), preserving theirs verbatim (P1, P3).

\emph{Speculative mode.} Write the schedule of \S\ref{sec:design:online} as a
sequence of states $x=x^{(0)}\!\to x^{(1)}\!\to\cdots\to x^{(s)}=\tilde{x}$,
where $s$ is the total number of steps. 
For a vertex at one specific step $i$, our algorithm will either keep the original value of the vertex as it is or issue a commit that changes its value. 
For any triangle T in the grid, at most one vertex is modified at one step due to the carefully designed parallel schedule (see Figure~\ref{fig:schedule}). 
If the original value is kept for a vertex, we have $\sigma(T;x^{(i)})=\sigma(T;x^{(i-1)})$ trivially for all its incident triangles, as nothing has been changed. 
If one commit is issued, $\sigma(T;x^{(i)})=\sigma(T;x^{(i-1)})$ is also ensured for all its incident triangles, as our verification admits the commit \emph{only if}
such condition holds (line 10 in Algorithm~\ref{alg:spec}). 
Every step therefore preserves every signature, and we have $\sigma(T;\tilde{x})=\sigma(T;x)$ for all $T$ by
induction on $i$. 
This guarantees that the decompressed field $\tilde{x}$ has the same signatures on all triangles as the original data $x$ and thus the same critical points.

Both guarantees are deterministic (zero lost, spurious, or type-changed
points), and the decompressor needs no side information beyond the committed
codes and raw specials.

%% ===================================================================
%% S4 IMPLEMENTATION  (~0.5-1 page)
%% ===================================================================
\section{Implementation}
\label{sec:impl}

We implement \sys in CUDA. For the lossless backend, we reuse mature
components: the Lorenzo prediction kernels and the GPU
Huffman codec from cuSZ~\cite{tian2020cusz, tian2021revisiting}, and the ANS coder from NVIDIA
nvCOMP\,5~\cite{nvcomp}. Our additions are
the topology-aware frontends, error-bound-aware extensions to the Lorenzo
kernels (per-element and per-tile bound variants), and the whole-pipeline
orchestration. Both modes default to the ANS coder. Fields are kept as structure-of-arrays: $U$ and $V$ occupy separate
row-major buffers and produce independent quantization and error-bound-ID
streams throughout.

\subsection{Block-Wise Mode}

The block-wise
compressor runs entirely on the device as four stages: (1) error-bound
derivation, (2) special classification with per-tile bound reduction, (3)
Lorenzo prediction and quantization under the per-tile uniform bound, and
(4) entropy coding. Stage~1 is a
$32\!\times\!16$-thread kernel that stages a vertex patch and its two-vertex halo
in shared memory 
% (padded \texttt{[TileY][TileX{+}1]} tiles to avoid bank conflicts),
($x$-padded to avoid bank conflicts), computes the per-triangle safe bound, and min-reduces it over each
vertex's incident cells. Stage~2 is a single \emph{fused} kernel that replaces
four earlier passes: it detects degenerate points ($U\!=\!V\!=\!0$) and zero-bound
(exact) vertices, rewrites their IDs, emits bit-packed masks with
\texttt{\_\_ballot\_sync}, and warp/block-min-reduces the IDs to one common
bound per $32\times32$ tile, all in a single sweep of the field, removing three
full-field re-reads and the intermediate arrays they need. The
tile-uniform bound pays off twice: it shrinks the error-bound side channel
from $n$ symbols to $n/1024$ (stored raw), and it lets the tile use a parallel
prefix-sum Lorenzo instead of the sequential row-scan variant, which we
measure at roughly $8\times$ faster. Exact
zero-bound values are gathered into side streams by population-counting the
bit-masks and a CUB exclusive scan~\cite{merrill2016scan}, so no per-element flag array is ever
materialized.

\subsection{Speculative Mode}

The speculative mode issues the 13-launch schedule 
(\S\ref{sec:design:online}) with 256 threads per block, 1 thread per target
vertex handling both components. The speculate--verify--back-off loop uses no
shared memory, reading neighbors from global memory over a
$\pm\text{step}$ stencil, and applies the two-tier signature test 
(\S\ref{sec:design:signature}). Raw and anchor values are
stream-compacted with a bit-mask plus a CUB exclusive scan before coding.
Critical-point predicates use double precision in this mode (the block-wise
mode's analytic bound runs in single precision), making the topological decision 
never a source of error itself.

\subsection{Entropy Backend and $U$/$V$ Concurrency}

\begin{figure}[htb]
  \centering
  \begin{tikzpicture}[x=0.62cm,y=0.62cm,
      cell/.style={draw=cprslate,thin,minimum width=6.2mm,minimum height=5mm,
                   inner sep=0.5pt,font=\footnotesize,anchor=center},
      zz/.style={cell,fill=cprslate!10},
      lowc/.style={cell,fill=cprbbg},
      highc/.style={cell,fill=cprbg},
      hdr/.style={font=\footnotesize,text=cprslate,anchor=south},
      ar/.style={->,>={Stealth[length=1.5mm]},cprslate!80,thin},
    ]
    % ---- left block: quantization codes (top) -> zig-zag (bottom) ----
    \node[hdr] at (2.5,0.42) {quant $q$};
    \foreach \v/\c in {4097/0,4095/1,4098/2,4096/3,4300/4,4094/5}
      {\node[cell] at (\c,0){\v};}
    \node[hdr] at (2.5,-1.18) {$q-r$, zig-zag};
    \foreach \v/\c in {2/0,1/1,4/2,0/3,408/4,3/5}
      {\node[zz] at (\c,-1.6){\v};}
    \draw[ar] (0,-0.45) -- (0,-1.15);              % quant -> zig-zag
    % ---- right block: low-byte plane (top), high-byte plane (bottom) ----
    \node[hdr] at (9.4,0.42) {low byte $\to$ ANS};
    \foreach \v/\c in {2/6.9,1/7.9,4/8.9,0/9.9,152/10.9,3/11.9}
      {\node[lowc] at (\c,0){\v};}
    \node[hdr] at (9.4,-1.18) {high byte $\to$ ANS $\approx 0$\,B};
    \foreach \v/\c in {0/6.9,0/7.9,0/8.9,0/9.9,1/10.9,0/11.9}
      {\node[highc] at (\c,-1.6){\v};}
    % zig-zag values split into the two byte planes
    \draw[ar] (5.55,-1.6) -- (6.35,-0.1);
    \draw[ar] (5.55,-1.6) -- (6.35,-1.6);
  \end{tikzpicture}
  % \vspace{-2mm}
  \caption{ANS frontend: a $16$-bit quantization code (radius $r$) is zig-zag
  mapped, then split into low- and high-byte planes coded independently.}
  \label{fig:byteplane}
  \Description{Diagram of the ANS frontend: a 16-bit quantization code is
  zig-zag mapped to a small unsigned integer, then split into a low-byte plane
  and a high-byte plane that is almost entirely zeros.}
\end{figure}
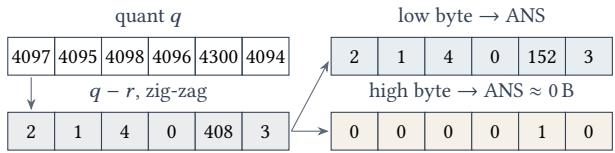

We drive both coders through thin bridges over the vendored kernels. Because
$U$ and $V$ are coded independently, these bridges overlap their work: the two Huffman
codebooks are built concurrently on separate host threads and the two
histograms run on separate CUDA streams, while the two encode passes are then
issued serially, since one parallel Huffman encode already saturates the GPU. The ANS path adds the byte-plane transform of Figure~\ref{fig:byteplane}
before coding: because most zigzag-mapped residuals are small, the high-byte
plane is almost entirely zeros and compresses to nearly nothing, which raises
the ratio. The four planes (the low and high bytes of $U$ and $V$) are
compressed together as a single four-plane nvCOMP ANS batch.

%% ===================================================================
%% S5 EVALUATION  (~3 pages)
%% ===================================================================
\section{Evaluation}
\label{sec:eval}

In this section, we evaluate \sys on three vector-field datasets, measuring its topology preservation, fidelity,
compression ratio, and throughput against state-of-the-art GPU and CPU
compressors on a GPU cluster~\cite{anoncluster}.

\subsection{Experimental Setup}
\label{sec:eval:setup}

\noindent\textbf{Platform.} All experiments run on a single NVIDIA
A100-SXM4 GPU in a dual-socket host with two AMD EPYC~7742 processors. We compile with
CUDA~12.6.

\noindent\textbf{Datasets.} We evaluate on three 2D vector fields, each with
two \texttt{float32} components $(u,v)$ per vertex on a triangular mesh
(Table~\ref{tab:datasets}). \emph{DT-10K} and \emph{DT-20K} are 2D decaying
turbulence fields generated with the Basilisk solver~\cite{basilisk} from
random band-limited initialization, at
$10000\times10000$ and $20000\times20000$ resolution. \emph{Ocean} is a
smaller $2400\times3600$ ocean-current field.
% [OPTIONAL: two additional larger datasets to be added.]

\begin{table}[t]
  \centering
  % \footnotesize = 8pt at the 10pt base: the smallest the CFP allows in tables.
  \footnotesize
  \caption{Vector-field datasets. Each vertex stores $(u,v)$ as
  \texttt{float32}, so the raw size is $\#\text{vertices}\times 8$\,bytes.}
  \label{tab:datasets}
  \begin{tabular}{@{} lrrrr @{}}
    \toprule
    Dataset & Resolution & \#Vertices & Raw size & \#Critical pts \\
    \midrule
    Ocean   & $2400\times3600$   & 8.64\,M  & 66\,MB   & $20{,}929$ \\
    DT-10K  & $10000\times10000$ & 100\,M   & 763\,MB  & $11{,}500$ \\
    DT-20K  & $20000\times20000$ & 400\,M   & 2.98\,GB & $228$ \\
    \bottomrule
  \end{tabular}
\end{table}

\noindent\textbf{Baselines.} We compare \sys against a broad set of GPU
error-bounded compressors:
cuSZ~\cite{tian2020cusz} and its interpolation-predictor variant
cuSZ-i~\cite{liu2024cuszi}, the ultra-fast cuSZp~\cite{huang2023cuszp}, and the
transform-based cuZFP~\cite{lindstrom2014zfp}.
We additionally compare against CPU cpSZ as the topology-preserving
reference~\cite{liang2023cpsz}. Only \sys and cpSZ preserve critical points; the
others are topology-agnostic and included to quantify the ratio, quality, and
throughput \sys trades for its guarantee.

\noindent\textbf{\sys\ configurations.} \sys runs in two modes, block-wise
(\sys-B) and speculative per-point (\sys-S), and each mode pairs with either a
Huffman (H) or an ANS (A) entropy coder, giving four configurations:
\sys-B-H, \sys-B-A, \sys-S-H, and \sys-S-A. Coding is lossless, so the coder
changes only ratio and throughput, never fidelity or topology. Unless a coder
is named we use the ANS coder for both, \sys-B-A and \sys-S-A, and write \sys-B
and \sys-S for these defaults.

\noindent\textbf{Metrics.} For quality we report rate-distortion
performance. The rate is the \emph{bit-rate}, $32/\text{compression ratio}$
for \texttt{float32} data. Distortion has two faces: we measure pointwise
fidelity by PSNR and topological fidelity by the P1/P2/P3 violation counts
(lost, spurious, and type-changed critical points) relative to the original
field. For performance we report \emph{compression} and \emph{decompression}
throughput as \emph{GPU end-to-end throughput}, the input size divided by the
wall-clock from the input field resident on the device to the output blobs
ready on the device.

% ============================ DATA STATUS ============================
% Have (REAL, 2026-07-25 CSVs): \sys-B (offline) & \sys-S (online), both codecs
% (hf/ans), ratio + compress/decompress throughput in GB/s (user-confirmed
% unit; the "gibs" column name is a misnomer), PSNR, SSIM, CP verify, tau in
% {0.1,0.075,0.05,0.025,0.01}, on DT-10K/DT-20K/Ocean. All throughput = END-TO-
% END.
% CONFIRMED (measured): \sys 0/0/0 CP violations (FTK verify_pass) across ALL
% 60 configs. SSIM: offline >0.99999, online >=0.9987. PSNR: offline 82-117 dB,
% online 53-91 dB. Platform = A100-SXM4-40GB. Throughput from the CLEAN CSVs
% (verify_ssim CSVs have noisy throughput outliers, e.g. ocean offline/ans 0.3
% GB/s -- do NOT use for throughput).
% PENDING: kernel time (CPR not run yet; for cuSZp kernel==E2E per user);
% baselines cuSZ, cuSZ-hi, cuSZ-i, cuZFP, cpSZ (cuSZp done). tab:timing stays
% [FILL] until CPR kernel times land. fig:rd = CPR curves now, baselines later.
% tab:tile REFRESHED on A100 (2026-07-26): offline/hf, eb=0.1, tile 8/16/32,
% ratio/PSNR/comp-GB-s; tile32 matches main tables (DT-10K 6.407/82.15).
% =====================================================================

\subsection{Topology Preservation and Fidelity}
\label{sec:eval:cp}

\begin{figure}[t]
  \centering
  \includegraphics[width=\columnwidth, trim={0 4ex 0 0}]{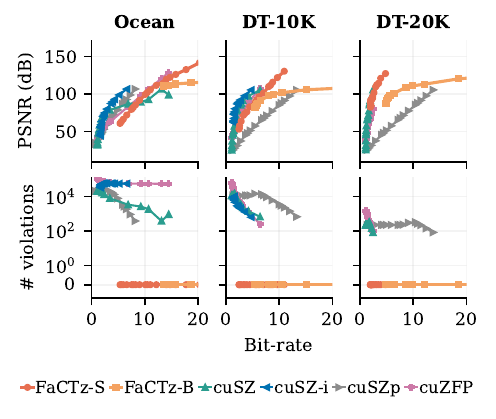}
  \caption{Fidelity and topology versus bit-rate, one column per dataset.
  \emph{Top:} PSNR. \emph{Bottom:} critical-point violations
  ($\text{P1}+\text{P2}+\text{P3}$).
  \sys-B and \sys-S hold at zero violations across all bit-rates and datasets;
  no baseline does. cuSZ-i is N/A on DT-20K. cuZFP is fixed-rate and is run at
  bit-rates matching cuSZ's. Bit-rate $=32/\text{ratio}$.}
  \label{fig:cp}
  \Description{Grid of line plots, one column per dataset: PSNR versus bit-rate
  in the top row and the number of critical-point violations versus bit-rate in
  the bottom row. The two \sys curves stay at zero violations across all
  bit-rates, while every baseline remains well above zero.}
\end{figure}

The central claim is that \sys preserves critical points that topology-agnostic
compressors sacrifice, and does so without giving up fidelity. We extract the
critical points of the original and reconstructed fields with
FTK~\cite{guo2021ftk} and count P1--P3 violations. Figure~\ref{fig:cp} plots, for each dataset, PSNR (top) and
the number of critical-point violations $\text{P1}+\text{P2}+\text{P3}$
(bottom) against bit-rate, for \sys and every GPU baseline.

\sys-B and \sys-S sit at \emph{exactly zero}
violations across the whole bit-rate range and every error bound, on all
datasets. No baseline does, and their failure splits in two ways. The
prediction-based cuSZ and cuSZp reduce topology error only by spending more
bits: their curves fall as bit-rate rises, but flatten out well above zero and
reach acceptable topology only near lossless, where the compression ratio is
negligible. The transform-based cuZFP and the interpolation-based cuSZ-i are worse
and, tellingly, almost \emph{bit-rate-independent}: they inject tens of
thousands of spurious critical points no matter how many bits they spend.
cuZFP still reports $\approx\!55{,}000$ violations on Ocean at $129$\,dB PSNR.
High pointwise fidelity therefore does not imply topological correctness, and
no baseline buys that correctness with more storage (O1).
The top row shows the converse is cheap for \sys: \sys-B is near-lossless (the
highest-PSNR curve on every dataset) and \sys-S stays within the fidelity band
of the strongest baselines, so the topological guarantee costs little PSNR.

\subsection{Comparison with GPU State-of-the-Art}
\label{sec:eval:throughput}

\S\ref{sec:eval:cp} compared all methods across the full rate range. We
now adopt a stricter and, for topology, more honest criterion: each method is
compared only at the error bound where it actually preserves every critical
point. At any looser bound the baseline corrupts topology
(\S\ref{sec:eval:cp}) and the comparison is meaningless; only at its
topology-preserving bound do the baseline and \sys deliver the same guarantee.
Table~\ref{tab:cpcmp} reports, for each baseline and dataset, the loosest such
bound, found by sweeping the relative and absolute bounds downward from
$10^{-5}$ in half-decade steps ($1$ and $5$ per decade) and finally testing the
limiting case $\mathrm{eb}=0$, and the
compression ratio there, against \sys's two modes, which preserve topology at
\emph{every} bound. cuZFP is omitted: its only CUDA mode is fixed-rate and
never preserves all critical points, and preservation is reachable only through
ZFP's CPU-only near-lossless and reversible modes.
The table also includes a GPU \emph{lossless} compressor, nvCOMP's
Zstd~\cite{nvcomp}, as the trivial end of the same spectrum: it preserves every critical point by
construction, so its ratio is the floor any topology-preserving lossy
compressor must beat to be worth using. That floor is low, $1.08$--$1.60\times$
at $3.9$--$4.5$\,GB/s compression, whereas \sys delivers the same guarantee at
$2.4$--$16.6\times$ and roughly an order of magnitude higher throughput.

The table makes the cost of the topological guarantee stark for the
baselines. On the DT-10K and Ocean fields, cuSZ and cuSZ-i never preserve
topology at any bound down to $0$: once the bound is tight enough to matter,
cuSZ misses every critical point in the field, and cuSZ-i's spline predictor
never yields a topology-preserving reconstruction. The methods that do preserve (cuSZp on every field, and cuSZ only on
DT-20K) require a
near-lossless bound, and their compression ratio there collapses to
$1.4$--$3.4\times$. \sys, in contrast, preserves every critical point while
still compressing substantially (\sys-B at $2.4$--$6.7\times$ and \sys-S at
$6.0$--$16.6\times$ across the three fields), a far higher ratio than any
baseline reaches for the identical guarantee, and entirely on the GPU.

\begin{table}[t]
  % \footnotesize = 8pt at the 10pt base: the smallest the CFP allows in tables.
  \centering\footnotesize
  \setlength{\tabcolsep}{4pt}
  \caption{Comparison at each method's topology-preserving bound: the loosest
  \emph{setting} and error bound giving zero critical-point violations, with the
  compression ratio (CR) and GPU end-to-end throughput there. \sys preserves
  topology at every bound.}
  \label{tab:cpcmp}
  \begin{tabular}{@{} l l l l r r r @{}}
    \toprule
    Dataset & Compressor & Set. & eb & CR
      & \shortstack{Comp.\\(GB/s)} & \shortstack{Decomp.\\(GB/s)} \\
    \midrule
    Ocean
      & cuSZ\textsuperscript{a}   & n/a & n/a & n/a & n/a & n/a \\
      & cuSZ-i\textsuperscript{a} & n/a & n/a & n/a & n/a & n/a \\
      & cuSZp  & rel & $5\mathrm{e}{-}9$ & 2.00$\times$ & 39.1 & 29.0 \\
      & nvCOMP Zstd\textsuperscript{c} & -- & -- & 1.60$\times$ & 3.9 & 5.7 \\
      & \sys-B & -- & -- & \textbf{2.36$\times$} & 36.0 & 63.9 \\
      & \sys-S & -- & -- & \textbf{6.01$\times$} & 29.4 & 72.8 \\
    \midrule
    DT-10K
      & cuSZ\textsuperscript{a}   & n/a & n/a & n/a & n/a & n/a \\
      & cuSZ-i\textsuperscript{a} & n/a & n/a & n/a & n/a & n/a \\
      & cuSZp  & rel & $1\mathrm{e}{-}8$ & 1.37$\times$ & 88.6 & 133.5 \\
      & nvCOMP Zstd\textsuperscript{c} & -- & -- & 1.08$\times$ & 4.3 & 28.6 \\
      & \sys-B & -- & -- & \textbf{5.98$\times$} & 47.0 & 133.2 \\
      & \sys-S & -- & -- & \textbf{12.99$\times$} & 45.2 & 85.8 \\
    \midrule
    DT-20K
      & cuSZ   & rel & $5\mathrm{e}{-}8$ & 3.36$\times$ & 58.3 & 42.9 \\
      & cuSZ-i\textsuperscript{b} & n/a & n/a & n/a & n/a & n/a \\
      & cuSZp  & rel & $5\mathrm{e}{-}8$ & 1.50$\times$ & 123.0 & 275.5 \\
      & nvCOMP Zstd\textsuperscript{c} & -- & -- & 1.09$\times$ & 4.5 & 35.8 \\
      & \sys-B & -- & -- & \textbf{6.67$\times$} & 60.1 & 139.2 \\
      & \sys-S & -- & -- & \textbf{16.63$\times$} & 41.5 & 89.4 \\
    \bottomrule
  \end{tabular}

  \smallskip
  \raggedright
  {\footnotesize
  \textsuperscript{a}Never preserves all critical points: we swept both absolute
  and relative bounds down to $5\times10^{-12}$ without recovering the full set
  (cuSZ-i crashed on these grids).\quad
  \textsuperscript{b}cuSZ-i's spline predictor cannot run on the
  $20000\!\times\!20000$ grid (a known upstream limit at large 2D sizes), so it is
  N/A there.\quad
  \textsuperscript{c}GPU \emph{lossless} compressor: it preserves topology by
  construction ($\mathrm{eb}=0$), so its ratio is the incompressible floor
  for this data.\par}
  % \vspace{-2mm}
\end{table}

\begin{figure}[t]
  \centering
  \includegraphics[width=.8\columnwidth, trim={0 2.5ex 0 0}]{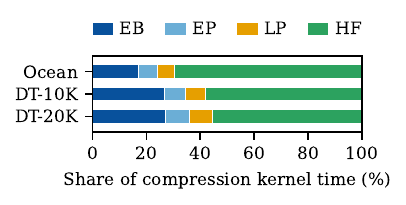}
  % \vspace{-2mm}
  \caption{Per-stage compression-kernel time of \sys-B-H (the Huffman path) at
  $\tau=0.1$, normalized per dataset. Blue stages (EB, EP) are
  critical-point-specific; the SZ backbone (LP, HF) is the rest.}
  \label{fig:breakdown}
  \Description{Horizontal 100-percent stacked bar chart of \sys-B compression
  kernel time per dataset, split into critical-point-specific stages (bound
  derivation and exception packing) and the SZ backbone (Lorenzo prediction
  and Huffman coding), the latter taking about 70 percent.}
\end{figure}

\begin{figure*}[tbh]
  \centering
  \includegraphics[width=\textwidth, trim={0 4ex 0 0}]{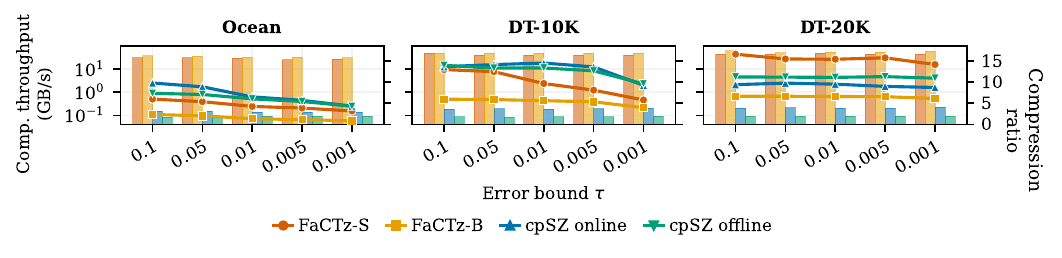}
  % \vspace{-1em}
  \caption{GPU versus CPU cpSZ across error bounds, one panel per dataset:
  compression throughput (grouped bars, left log axis) and compression ratio
  (lines, right axis) for \sys-S, \sys-B, and our OpenMP-parallel CPU cpSZ in
  its online and offline modes.}
  \label{fig:cpsz}
  \Description{Three panels, one per dataset, with grouped bars for compression
  throughput on a logarithmic axis and lines for compression ratio, comparing
  \sys-S, \sys-B, and OpenMP cpSZ in its online and offline modes across five
  error bounds.}
\end{figure*}

\noindent\textbf{Throughput in context.} \sys is slower than the topology-agnostic
GPU compressors at any \emph{fixed} bound, which is expected as it performs work
they omit entirely: every candidate reconstruction is checked against the
critical-point signatures of its incident triangles, and the block-wise mode
additionally derives an analytic safe bound for every triangle. But a fair speed reference must also preserve topology, and no GPU baseline
does so at a useful ratio (\S\ref{sec:eval:cpsz}).
A per-stage breakdown of \sys-B (Figure~\ref{fig:breakdown}) localizes the
cost of the guarantee. The critical-point-specific work, bound derivation (EB) and the
exception-packing kernels (EP) that follow it to classify and compact the
vertices needing exact ($\mathrm{eb}=0$) treatment, accounts for only about
$30\%$ of the compression kernel time; the remaining $\sim\!70\%$ is the
conventional SZ backbone of Lorenzo prediction (LP) and Huffman coding (HF),
itself dominated by entropy coding. The topological
guarantee is thus a bounded surcharge on a standard prediction-plus-coding
pipeline rather than a pervasive overhead, and the single largest cost is the
shared entropy coder, which ANS alone accelerates substantially
(\S\ref{sec:eval:cfgthrpt}). cuSZp avoids that cost altogether: its
fixed-length encoder skips variable-length entropy coding, which is why it is
faster and why it compresses so little.

\subsection{Comparison with CPU cpSZ}
\label{sec:eval:cpsz}

As \S\ref{sec:eval:cp} established, the only existing compressor that preserves
critical points is the CPU-based cpSZ, so comparing \sys's GPU throughput
against a CPU-serial cpSZ would be unfair. To make the comparison fair, we implement an OpenMP-parallel version of cpSZ ourselves. Its
offline (two-pass) variant parallelizes the per-triangle bound derivation over
even and then odd grid rows, which is race-free because each vertex's bound is
a pointwise minimum and hence order-independent. Its online (single-pass)
variant is inherently sequential (each cell depends on its up, left, and
upper-left neighbors), so we parallelize it as a \emph{wavefront}: cells on the
same anti-diagonal are mutually independent and run concurrently, with a
barrier between diagonals preserving the serial dependency order.

Figure~\ref{fig:cpsz} compares both \sys\ modes against this parallel cpSZ,
run with $128$ threads on a separate AMD EPYC~7713 CPU, across five error
bounds spanning $10^{-1}$ to $10^{-3}$ on all three datasets. In throughput
(bars, left log
axis), both \sys\ modes are roughly two orders of magnitude faster: \sys\
sustains $25$--$60$\,GB/s while even the $128$-thread cpSZ reaches only
$\sim\!0.1$--$0.2$\,GB/s, a $174$--$643\times$ speedup mode-for-mode. In
compression ratio (lines, right axis), cpSZ's two variants reach
$4$--$15\times$: its framework stacks the Zstd lossless
compressor~\cite{zstd, wu2025orchestration} on top of
entropy coding, which raises the ratio but further lowers its throughput. This
puts cpSZ above \sys-B on every field and above \sys-S on
Ocean, whereas \sys-S stays comparable on DT-10K ($13.0$ versus $13.7\times$
at $\tau{=}0.1$) and clearly higher on DT-20K ($16.6$ versus $9.4\times$). \sys\ thus turns topology-preserving compression
from a pipeline bottleneck (O2) into a negligible cost, and its speculative
mode keeps the ratio competitive with the CPU reference while doing so.

\subsection{Throughput across Configurations}
\label{sec:eval:cfgthrpt}

\begin{figure*}[!t]
  \centering
  \includegraphics[width=1.0\linewidth]{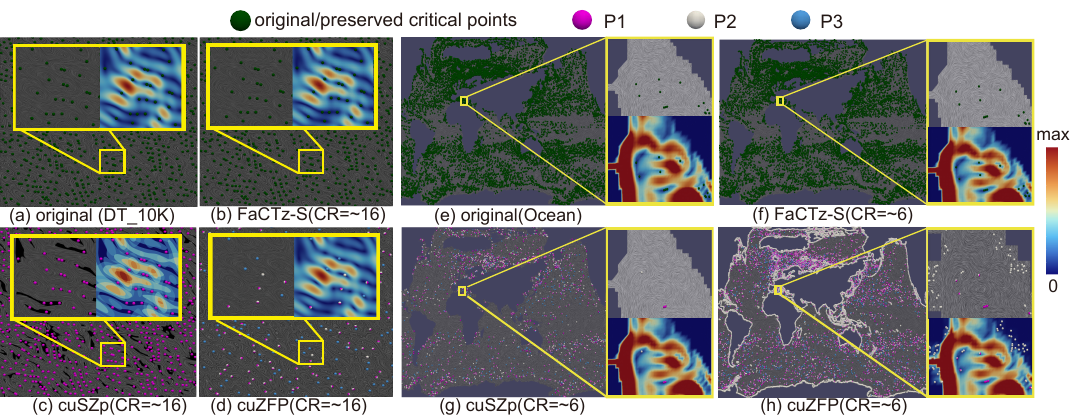}
  % \vspace{-3mm}
  \caption{Critical points overlaid on reconstructions at matched compression
  ratio, with LIC visualized as context: DT-10K turbulence (left, $\mathrm{CR}\approx16$, first
  $2500\times2500$ region) and Ocean (right, $\mathrm{CR}\approx6$). Within each
  field: (a,\,e)~original, (b,\,f)~\sys-S, (c,\,g)~cuSZp, (d,\,h)~cuZFP. Points are
  colored by fate---green: preserved; magenta, gray, blue: lost (P1), spurious
  (P2), and type-changed (P3). \sys-S preserves every point (all green); the
  topology-agnostic baselines scatter P1--P3 violations. Insets zoom the boxed
  region; background color is vector-field magnitude.}
  \label{fig:vis_2_dataset}
  \Description{Vector-field magnitude images with extracted critical points
  overlaid and colored by fate, comparing the original with \sys-S, cuSZp, and
  cuZFP reconstructions on the DT-10K and Ocean fields. \sys-S points are all
  green (preserved), while the baselines scatter lost, spurious, and
  type-changed points.}
\end{figure*}

\begin{figure}[htb]
  \centering
  \includegraphics[width=\columnwidth, trim={0 4ex 0 0}]{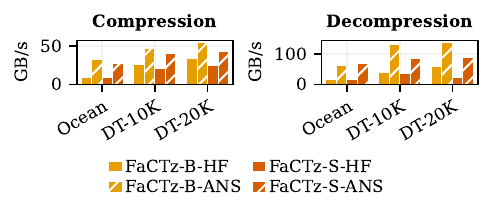}
  % \caption{Per-configuration GPU end-to-end throughput, averaged over
  % $\tau\in\{0.1,\dots,0.01\}$.Compression (left) and decompression (right) for the four \sys\ configurations
  % across the three datasets. Color denotes the mode (\sys-B / \sys-S); hatching
  % denotes the ANS coder.}
  \caption{Per-configuration GPU end-to-end throughput.}
  \label{fig:thrpt_cfg}
  \Description{Two grouped bar charts of GPU end-to-end throughput, compression
  on the left and decompression on the right, for the four \sys configurations
  across the three datasets.}
\end{figure}

Figure~\ref{fig:thrpt_cfg} breaks throughput down across the four \sys\
configurations, and three trends hold on every dataset. First, the ANS coder is
faster than Huffman in both stages, and markedly so on decompression:
\sys-B-ANS decompresses DT-10K at $131$\,GB/s against $41$\,GB/s for \sys-B-HF.
Second, the block-wise mode is generally faster than the speculative one,
especially on compression: the per-point speculate--verify--back-off loop
performs strictly more work, the price \sys-S pays for its higher compression
ratio (\S\ref{sec:eval:cpsz}). Third, decompression is faster than compression
throughout (often by $2$--$3\times$ with ANS), because decoding merely replays
the recorded decisions, with no bound derivation or verification. In absolute
terms \sys\ compresses at up to $\sim\!60$\,GB/s and decompresses at up to
$\sim\!140$\,GB/s, faster than the interpolation-based cuSZ-i
($5$--$15$\,GB/s) while additionally preserving every critical point.

% ===== \S5.6 Sensitivity: Lorenzo Tile Size --- commented out per request (kept for later) =====
% \subsection{Sensitivity: Lorenzo Tile Size}
% \label{sec:eval:sensitivity}
%
% The block-wise mode's tile size trades compression ratio, fidelity, and
% throughput. Table~\ref{tab:tile} sweeps tile sizes $8$, $16$, and $32$ at
% $\tau=0.1$. A larger tile consistently raises both PSNR and throughput, since
% one common bound covers more vertices with fewer per-vertex decisions; the
% ratio, however, moves in a dataset-dependent way---rising with tile size on
% DT-20K, peaking at tile $16$ on DT-10K, and falling on Ocean, where the larger
% tile trades ratio for a steep fidelity gain ($81$ to $111$\,dB). Tile $32$
% gives the best fidelity and throughput on every dataset, so we adopt it as the
% default, accepting a slightly lower ratio on DT-10K and Ocean; critical points
% are preserved (0/0/0) at every tile size.
%
% \begin{table}[t]
%   \centering\small
%   \caption{Block-wise sensitivity to Lorenzo tile size at $\tau=0.1$ (offline
%   mode, Huffman coder, A100). Each cell: compression ratio / PSNR (dB) /
%   compression throughput (GB/s). All configurations preserve critical points
%   (0/0/0).}
%   \label{tab:tile}
%   \begin{tabular}{lccc}
%     \toprule
%     Dataset & Tile 8 & Tile 16 & Tile 32 \\
%     \midrule
%     DT-10K & 6.64 / 71.1 / 26.0 & 6.94 / 75.8 / 30.9 & 6.41 / 82.1 / 33.5 \\
%     DT-20K & 5.35 / 80.9 / 35.7 & 6.34 / 83.9 / 43.2 & 6.89 / 87.0 / 44.1 \\
%     Ocean  & 3.65 / 81.3 / 9.9  & 3.09 / 94.0 / 10.3 & 2.45 / 110.6 / 10.6 \\
%     \bottomrule
%   \end{tabular}
% \end{table}

\subsection{Visual Comparison of Reconstructions}
\label{sec:eval:vis}

Figure~\ref{fig:vis_2_dataset} makes the contrast visible on DT-10K
($\mathrm{CR}\!\approx\!16$) and Ocean ($\mathrm{CR}\!\approx\!6$). On both
fields every reconstruction is visually near-indistinguishable from the
original, with fine-scale texture intact in the zoom insets. The critical
points, however, differ sharply: \sys-S is entirely green (every critical
point preserved), whereas cuSZp and cuZFP scatter P1--P3 violations across the
domain, most densely in the turbulent regions where critical points
concentrate.

% \begin{figure*}[t]
%   \centering
%   \includegraphics[width=\textwidth]{figures/ocean_v1.pdf}
%   \caption{\textcolor{red}{caption input}.}
%   \label{fig:vis_ocean}
% \end{figure*}

%% ===================================================================
%% S6 RELATED WORK  (~0.75-1 page)
%% ===================================================================
\section{Related Work}
\label{sec:related}

\noindent\textbf{Error-bounded lossy compression.} \emph{Prediction-based}
methods such as the SZ compressor family~\cite{di2016sz, liang2018error, zhao2021optimizing, liu2024high} and FPZIP~\cite{lindstrom2006fpzip} predict each value
from decoded neighbors and quantize the residual; \emph{transform-based}
methods (ZFP~\cite{lindstrom2014zfp}, SPERR~\cite{li2023sperr},
TTHRESH~\cite{ballester2020tthresh}, MGARD~\cite{ainsworth2019mgard}) truncate
transform coefficients. Both bound only the \emph{magnitude} of per-value error
and ignore derived topology, the gap \sys addresses~\cite{di2025survey}.

\noindent\textbf{GPU compressors.} To keep pace with data-generation rates,
these pipelines have been ported to GPUs. Representative compressors include cuSZ~\cite{tian2020cusz,
tian2021optimizing}, cuSZp~\cite{huang2023cuszp, huang2024cuszp2}, cuSZ-i~\cite{liu2024cuszi}, SZx~\cite{yu2022szx}, cuZFP~\cite{lindstrom2014zfp}, and LC~\cite{fallin2025lc}; the same
pipelines have also been carried to wafer-scale
engines~\cite{song2024ceresz, song2025wafer}. Together they reach tens to
hundreds of GB/s. \sys reuses such a backend (\S\ref{sec:impl}) but differs in kind:
it is the first GPU compressor to guarantee \emph{critical-point preservation}, rather than
merely enforcing a pointwise error bound.
% to enforce a \emph{topological} constraint during GPU
% compression, not merely a pointwise one.

\noindent\textbf{Topology-preserving compression.} A separate line preserves
derived topology under an error bound. For \emph{vector fields}, cpSZ preserves
critical points~\cite{liang2020toward, liang2023cpsz}, and its variation~\cite{xia2024sod} leverages sign-of-determinant predicates to make the
guarantee robust to numerical degeneracy. TspSZ~\cite{xia2025tspsz} further extends
preservation to the full topological skeleton. For
\emph{scalar fields}, methods preserve contour trees~\cite{yan2023toposz} and
Morse--Smale complexes~\cite{li2026msc, li2026pmsz}, a general framework
augments existing lossy compressors with topological
guarantees~\cite{gorski2025framework}, and a concurrent effort
reaches GPU and distributed scale but for scalar-field descriptors rather than
vector-field critical points~\cite{li2026exactz}. TFZ carries the idea to
second-order \emph{tensor fields}~\cite{gorski2026tfz}. The vector-field methods,
however, run on the CPU far below GPU rates (\S\ref{sec:bg:motivation}), and the
scalar-field efforts do not address vector-field topology.

%% ===================================================================
%% S7 CONCLUSION  (~0.4 page; 3 sentences)
%% ===================================================================
\section{Conclusion}
\label{sec:conclusion}

Error-bounded lossy compression is indispensable for GPU-scale scientific
simulation, yet a pointwise bound does not protect vector-field topology, and
the only compressor that does runs on the CPU far too slowly. We presented
\sys, the first GPU error-bounded lossy compressor that provably preserves
critical points, by decoupling the inherently coupled preservation constraint
into independent parallel work in two modes: an analytic block-wise mode for
throughput and a speculative per-point mode for ratio. \sys preserves every
critical point across all datasets and bounds while compressing on the GPU at
up to $60$\,GB/s, two orders of magnitude (up to $\sim\!640\times$) faster
than parallel CPU cpSZ, while its speculative mode roughly doubles the
block-wise ratio.

%% ===================================================================

\section*{Acknowledgments}
%% CFP requires generative-AI use to be disclosed in the Acknowledgments.
%% Plain \section* (not \begin{acks}) because acmart's anonymous mode drops
%% the acks environment entirely; switch to \begin{acks} for camera-ready.
% Generative AI tools assisted with language editing, figure generation,
% and \LaTeX{} formatting; all technical content, algorithms, and claims
% were authored and verified by the authors.
Generative AI tools were used to assist with language editing, figure-generation
scripts, and \LaTeX{} formatting. All technical content, algorithms, data, and
claims were authored and verified by the authors.

\bibliographystyle{ACM-Reference-Format}
\bibliography{cpr}

\end{document}